\documentclass[]{elsart3p}

\usepackage{graphicx,amscd,amsmath,amssymb}

\begin{document}

\journal{Physica A}

\begin{frontmatter}

\title{Universality in Bibliometrics}

\author[IFUFRGS]{Roberto da Silva\corauthref{cor}}, 
\ead{rdasilva@inf.ufrgs.br}
\author[IIUFRGS]{Fahad Kalil},
\ead{fahad@inf.ufrgs.br}
\author[Filo,INCT]{Alexandre Souto Martinez}  and
\ead{asmartinez@ffclrp.usp.br}
\author{Jos\'e Palazzo Moreira de Oliveira} 
\ead{palazzo@inf.ufrgs.br}

\corauth[cor]{Corresponding author.}

\address[Filo]{ Faculdade de Filosofia, Ci\^encias e Letras de Ribeir\~ao Preto (FFCLRP), \\
             Universidade de S\~ao Paulo (USP) \\ 
             Avenida Bandeirantes, 3900 \\ 
             14040-901, Ribeir\~ao Preto, S\~ao Paulo, Brazil.}

\address[IFUFRGS]{ Instituto de F\'{\i}sica\\
Universidade Federal do Rio Grande do Sul\\
Av. Bento Gon\c{c}alves 9500 \\ 
Caixa Postal 15051\\
 91501-970, Porto Alegre RS, Brazil}

\address[IIUFRGS]{ Instituto de  Inform\'atica\\
Universidade Federal do Rio Grande do Sul\\
Av. Bento Gon\c{c}alves 9500 \\ 
Caixa Postal 15051\\
 91501-970, Porto Alegre RS, Brazil}

\address[INCT]{Instituto Nacional de Ci\^encia e Tecnologia em Sistemas Complexos}%

\begin{abstract}
{
Many discussions have enlarged the literature in Bibliometrics since the Hirsh proposal, the so called $h$-index. 
Ranking papers according to their citations, this index quantifies a researcher only by its greatest possible number of papers that are  cited at least $h$ times.   
A closed  formula for $h$-index distribution that can be applied for distinct databases is not yet known. 
In fact, to obtain such distribution, the knowledge of citation distribution of the authors and its specificities are required.
Instead of dealing with researchers randomly chosen, here we address different groups based on distinct databases.  
The first group is composed by physicists and biologists, with data extracted from Institute of Scientific Information (ISI). 
The second group composed by computer scientists, which data were extracted from Google-Scholar system. 
In this paper, we obtain a general formula for the $h$-index probability density function (pdf) for groups of authors by using generalized exponentials in the context of escort probability. 
Our analysis includes the use of several statistical methods to estimate the necessary parameters. 
Also an exhaustive comparison among the possible candidate distributions are used to describe the way the citations are distributed among authors. 
The $h$-index pdf should be used to classify groups of researchers from a quantitative point of view,  which is meaningfully interesting to eliminate obscure qualitative methods.  
}
%
%
\end{abstract}

\end{frontmatter}



\section{Introduction}

\label{introduction}

The scientific community has not been the same after the publication of the
polemic index of Hirsh ($h$-index). If an author has $h$-index equal to $h$
means that she has $h$ papers with at least $h$ citations but she has not $%
h+1$ papers with at least $h+1$ citations \cite{Hirsch2005}. This definition
leads to an important fact: an author with index $h$ has at least $h^{2}$
citations \cite{Hirsch2005}, a lower bound in the citation number.

This simple but powerfull idea has also suscitated other informetric
formulations \cite{Egghe2006}. This index joins attributes as: productivity,
quality and homogeneity in the same measure. It has also been used as one of
the most important measures to quantify scientists in order to obtain fairer
rankings \cite{Ball2005}. Fairer rankings should mean, for example, grants
fairer distributed among scientists really based on capability therefore it
must stimulate a healthy competition.

Other successful variations of the $h$-index have been proposed. For
instance, dividing the $h$ top papers index by the average of authors in
these $h$ papers leads to: $h_{I}=h^{2}/N_{t}$~\cite{Batista2006}.
Considering that a massive part of publications cannot be used to compute
the $h$-index, an interesting and alternative index that considers the
weight of this mass of these lazy papers has been proposed in Refs~\cite%
{Egghe2006-II,Tol2008,Woeginger2008}). Other metrics consider that not an
individual $h$-index for each scientist might be considered but also its
version for a group of researchers, which is denoted as successive $h$-index~%
\cite{Egghe2008,Schubert2007}.

Laeherre and Sornette \cite{Laherrere1998} were the first to address the
researcher citation distribution. They have ranked 1120 physicists according
to their total number of citations. The number of researchers $N(x)$ as
function of their citation number $x$ follows a stretched exponential
function: 
\begin{equation}
N(x)=N_{0}\exp [-(x/x_{0})^{\beta }]  \label{Laherrere_intro}
\end{equation}%
with $\beta \approx 0.3$. Here $N_{0}=N(0)$ is number of authors with no
citation and $x_{0}$ an parameter that can be estimated for example if the
citation mean is known. Of course, citation number $x$ in an integer
variable, but here we have considered it as a continuous variable.

Alternatively, Redner \cite{Redner1998} has also addressed this questioning
in a slightly different way. The probability distribution of citations of
783.339 scientific papers, not authors, published in 1981, with the
6.716.198 citations obtained between 1981 and 1997 in the base Institute of
Scientific Information (ISI) has been studied. The envelope of this
distribution presents a stretched exponential behavior for low citation
number $x<x_{c}$, with $x_c = 200$ and for large citation number $x>x_{c}$,
the power law behavior is dominant $N(x)\sim x^{-\alpha }$, with $\alpha
\approx 3.0$.

Tsallis and Albuquerque \cite{Tsallis2000} observed that Redner's
probability distribution function (pdf) and the paper (not author) citation
pdf could be better described by a generalized exponential distribution that
covered the two situations ($x\leq x_{c}$ and $x>x_{c}$): 
\begin{equation}
N_{q}(x)=N_{0}[\exp _{q}(-\lambda x)]^{q}  \label{Tsallis_intro}
\end{equation}
where generalized exponential function ($q$-exponential) is $%
\exp_{q}(x)=[1+(1-q)x]^{1/(1-q)}$, if $(1-q)x\geq -1$ and it vanishes
otherwise. The $q$-exponential inverse function is the $q$-logarithm: $%
\ln_{q}(x)=(x^{1-q}-1)/(1-q)$~\cite{Tsallis1999,Tsallis2009,arruda_2008}).
The parameter $\lambda $ in Eq.~(\ref{Tsallis_intro}) is obtained
constraining the citation per paper average number to be a constant, $%
\left\langle x\right\rangle =\int_{0}^{\infty }dx\ x\ P_{q}(x)=$ constant,
where $P_{q}(x)=[\exp _{q}(-\lambda x)]^{q}/\int_{0}^{\infty }dy[\exp
_{q}(-\lambda y)]^{q}$. Although in Ref.~\cite{Tsallis2000}, this approach
has been used only for distribution of scientific papers, we show that it
can be also applied for author citation probability distribution.

Here, we consider the stretched and generalized exponential pdf's to
describe the envelope of the author citation distribution. Also, we ask
which pdf is more appropriate to describe the $h$-index distribution of
authors of distinct groups of researchers. Using a continuous approximation
for citation distributions, two different researcher groups are collected
from two different database. One from Graduate Programs in Physics and
Biology of public universities in Brazil using their ISI publication
registry. The other, from software engineering area, where we computed the
$h$-indices and the total citations of the members of program committees of
different conferences. It is important to stress, that our purpose, is to
show that the same $h$-index pdf is verified even in \textquotedblleft
soft\textquotedblright\ databases more suitable to areas that are not based
strictly on journal publications. If different models follow the same law,
in statistical physics, one says that this law is universal, in fact there
are classes of universality. We question, if the two mentioned groups of
authors, of different areas with data collected from different databases,
have the same pdf, and which one is better to describe the data.

This paper is organized as follows.  
In Sec.~\ref{sec2}, we show the details of deduction of the generalized distribution of $h$-indices extracted from different citation
distributions for different approaches: i) The first one establishes that
citation distribution follows a escort probability distribution according to
Eq.~(\ref{Tsallis_intro}); ii) The other one prescribes that citations are
distributed according to a stretched exponential via equation (\ref%
{Laherrere_intro}). 
In Sec.~\ref{databases} we present the details about databases used to test h-index distribution formula. We describe with some details how the data were extracted for each studied group.
In Sec.~\ref{results}, we present our results, which can be separated in two distinct parts. In a first (preliminary) part we
estimate necessary parameters of the citation distributions i) and ii) previously reported by using the method of moments and by comparisons of the
theoretical and empirical cumulated citation distribution. 
Secondly, we so obtain the $h$-index distribution by studying it for the distinct areas. 
The parameters obtained from $h$-index distribution fits are compared with the same parameters estimated by citation distribution fits and we conclude that generalized statistics produce a $h$-index distribution that corroborates the citation distribution differently from stretched exponential that points out more accentuated differences between these two ways of estimation. 
We finally summarize our results as well as highlight our main contributions in Sec.~\ref{conclusions}.

\section{$h$-index probability density functions for groups of authors}
\label{sec2}

In this section, we consider the continuous limit of normalized citation
distribution and we describe a deduction of the distribution of $h$-indices
for a group of authors for the stretched~\cite{Laherrere1998} and
generalized~\cite{Tsallis2000} exponential pdf's, which are confronted next
section.

The fundamental Hirsch hypothesis~\cite{Hirsch2005} leads to $x =c h^{2}$,
where $c$ is a constant, which is determined making a suitable linear fit.
Since the databases supplies the number of citations ($x_{i}$) and the
corresponding $h$-index of the $i^{\mbox{th}}$ author ($h_{i}$), with $i=1,
\ldots,n$, one has an estimator to mean citation number $\langle x \rangle$
and $c$. Collecting a sample of citations of all authors of a group, which
we denote by $x_{1},x_{2}, \ldots, x_{n}$ corresponding to the respective $n$
authors, the estimate of $\left\langle x\right\rangle $, denoted by $%
\widehat{x}$, is the simple arithmetic mean: 
\begin{equation}
\widehat{x}=\frac{x_{1}+x_{2}+...+x_{n}}{n}\;.
\end{equation}

Similarly, an estimator $\widehat{c}$ for the coefficient $c$ comes from
least square fitting: 
\begin{equation}
\widehat{c}=\frac{\sum\limits_{i=1}^{n}h_{i}^{2}x_{i}-\dfrac{1}{n}%
\sum\limits_{i=1}^{n}h_{i}^{2}\sum\limits_{j=1}^{n}x_{j}}{%
\sum\limits_{i=1}^{n}h_{i}^{4}-\dfrac{1}{n}\left(
\sum\limits_{i=1}^{n}h_{i}^{2}\right) ^{2}}\;,  \label{slope}
\end{equation}%
that measures the slope in the linear fit of the plot $x$ versus $h^{2}$.

\subsection{Stretched Exponential PDF}

One interesting pdf used to study the citation distribution is the stretched
exponential pdf tha we define in the continuous case as: 
\begin{equation}
P_{\beta }(x)=\frac{\beta }{x_{0}\Gamma (1/\beta )}\;e^{-(x/x_{0})^{\beta
}}\;.  \label{laeherre}
\end{equation}%
One can estimate $x_{0}$ by calculating $\left\langle x\right\rangle $,
which, in turn, is estimated by $\widehat{x}$, i.e., we can demand the
condition $\left\langle x\right\rangle =\int\limits_{0}^{\infty }dxxP_{\beta
}(x)=x_{0}\Gamma (2/\beta )/\Gamma (1/\beta )=\widehat{x}$, resulting in: 
\begin{equation}
\widehat{x}_{0}\approx \frac{\Gamma (1/\beta )}{\Gamma (2/\beta )}\;\widehat{%
x}\;.
\end{equation}%
According to $x=ch^{2}$, the $h$-index pdf for the stretched exponential is: 
\begin{equation}
H_{\beta }(h)=\frac{2\widehat{c}\beta \Gamma (2/\beta )}{\widehat{x}\Gamma
(1/\beta )^{2}}\;h\;\exp \left[ -\left( \frac{\widehat{c}\;\Gamma (2/\beta
)h^{2}}{\Gamma (1/\beta )\widehat{x}}\right) ^{\beta }\right] \;.
\label{h_distribution_by_sornete}
\end{equation}

In section \ref{results} we will show fits of the $h$-index histograms for the
distinct analyzed groups using this pdf. Now let us obtain another formula
for $h$-index pdf by using the proposal of generalized exponentials.    

\subsection{Generalized Exponential PDF}

The generalized exponential approach, considers: 
\begin{equation}
P_{q}(x)=\frac{[\exp _{q}(-\lambda x)]^{q}}{\int_{0}^{\infty
}dy[\exp_{q}(-\lambda y)]^{q}} \;,  \label{equation_citations_tsallis}
\end{equation}%
for $0<x<\infty $. To estimate $\lambda$, one analytically calculates the
first moment of this pdf $\left\langle x\right\rangle =\int_{0}^{\infty
}xP_{q}(x)dx=1/[(2-q)\lambda ]$, with does not diverge for $1<q<2$. Next,
one estimates $\lambda $ through 
\begin{equation}
\widehat{\lambda }\approx \frac{1}{(2-q)\widehat{x}}
\end{equation}%
and a hybrid expression for the citation pdf, considering that $\widehat{x}$
is an estimator for $\left\langle x\right\rangle $ is: 
\begin{equation}
\widehat{P}_{q}(x)=\frac{1}{(2-q)\widehat{x}}\;[\exp _{q}\{-x/[(2-q)\widehat{%
x}]\}]^{q}\;.  \label{citation_generalized}
\end{equation}%
Now we are able to compute the $h$-index distribution for a group of
researchers: 
\begin{equation}
\begin{array}{lll}
H_{q}(h) & = & \left\vert \frac{dx}{dh}\right\vert \;\widehat{\lambda }%
\;[\exp _{q}(-\widehat{\lambda }\widehat{c}h^{2})]^{q} \\ 
&  &  \\ 
& = & \dfrac{2\widehat{c}}{(2-q)\widehat{x}}\ h\ [\exp _{q}\left\{ -\frac{%
\widehat{c}h^{2}}{(2-q)\widehat{x}}\right\} ]^{q}\;,%
\end{array}
\label{h-index_distribution}
\end{equation}%
which is normalized. Since the parameters $\left\langle x\right\rangle $ and 
$c$ are estimated by $\widehat{x}$ and $\widehat{c}$, respectively, the only
free parameter is $q$.

Now we have two candidates for $h$-index pdf. These pdf's are used to fit
our data for the two different groups of researchers.

\section{Databases}
\label{databases}

Two different researcher groups are collected from two different databases.
The first database is obtained from 1203 researchers from 19 Graduate
Programs in Physics (600 researchers) and 26 Graduate Programs in Biology
(603 researchers) of public universities in Brazil. To obtain the published
papers, we have used the data from the research digital curriculum vitae at
public disposal at the Lattes database\footnote{%
The Lattes database provides high-quality data of about 1.6 million
researchers and about 4,000 institutions.} (http://lattes.cnpq.br/english),
which is a well-established and conceptualized database~\cite{Lane2010},
where most Brazilian researchers deposit their vitae. By a manual process,
each researcher publication has been extracted from this platform and
compared to the ISI-JCR database\footnote{%
http://apps.isiknowledge.com/} by crossing the information between these two
bases.

It is important to notice that refinements on queries were performed for suitably computing the $h$-index of author. 
We tune the details of query performed on ISI-JCR until number of papers of queried author in this database is the same or as close as possible of that one registered by author in its Lattes vitae. 
Only after this process, we compute the authors $h$-index.  

The same process is run out for each researcher of the studied group. 
This method, although manual is an excellent filter to obtain data for Brazilian researchers.

In many areas as Computer Science, the authors are ranked considering not only papers published in scientific journals, but also by papers published in important conferences. 
For sake of the simplicity, we concentrate our analysis in software engineering area by computing the $h$-indices and the total citations of the members of program committees of seven different conferences in a total of 600 researchers. 
For these authors, we have used the Harzing program, that computes the $h$-index based on google-scholar index\footnote{http://www.harzing.com/pop.htm}. 
The choice to use Google Scholar instead of ISI-JCR for computer scientists is because many important conferences are not captured by ISI-JCR database. 
It is important to stress, that our intention, is to show that the universality of $h$-index pdf is verified even in ``soft'' databases more suitable to areas that are not based strictly on journal publications. 
Also, the number of considered researchers in Harzing has not been greater because there is a blockage of the system after a number of searches, which make our job much more vagarious.

\section{Results}
\label{results}

\begin{figure}[h]
\begin{center}
\includegraphics[width=.8\columnwidth]{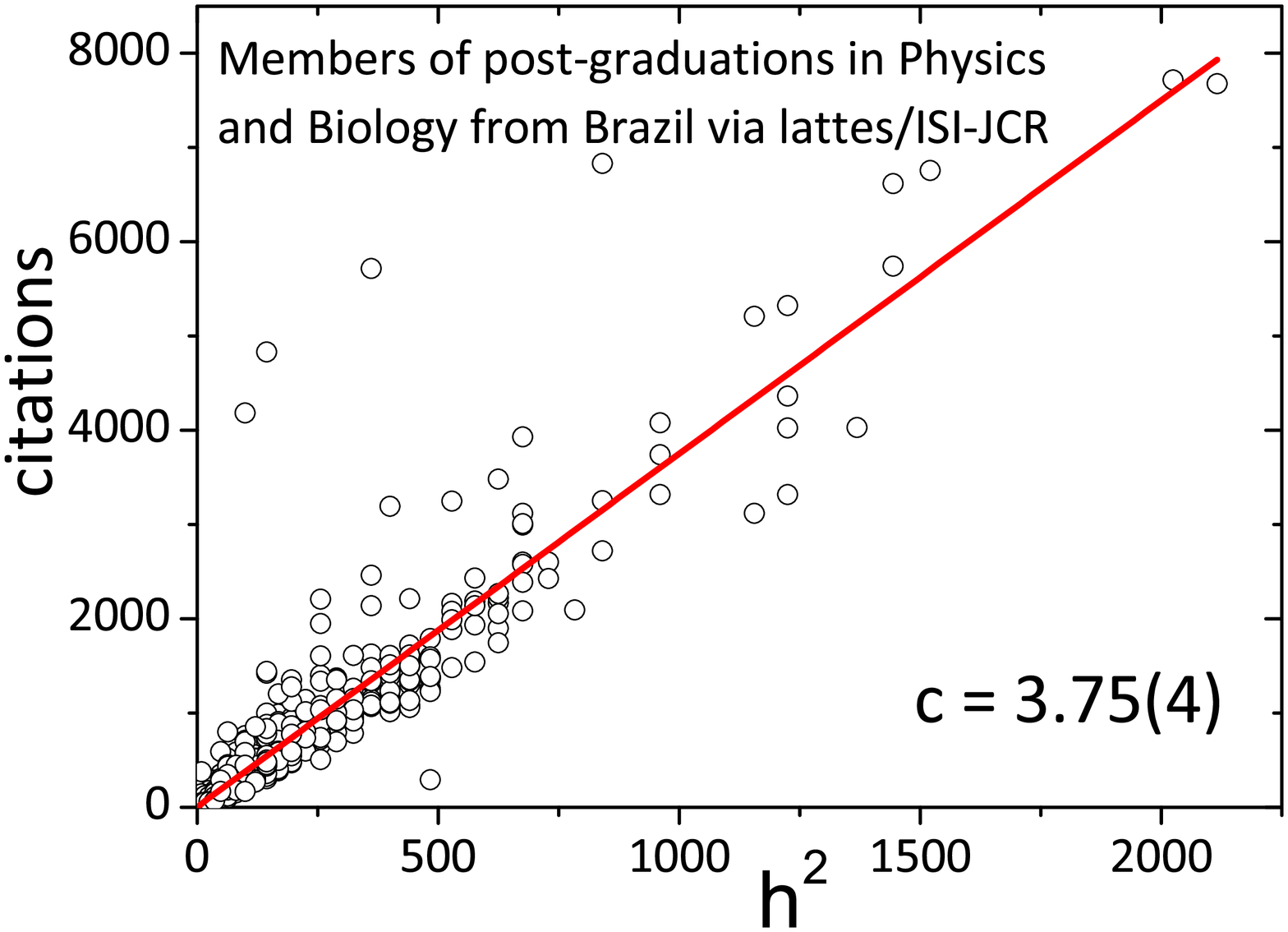} 
\\[0pt]
\textbf{(a)} \\[0pt]
\includegraphics[width=.8\columnwidth]{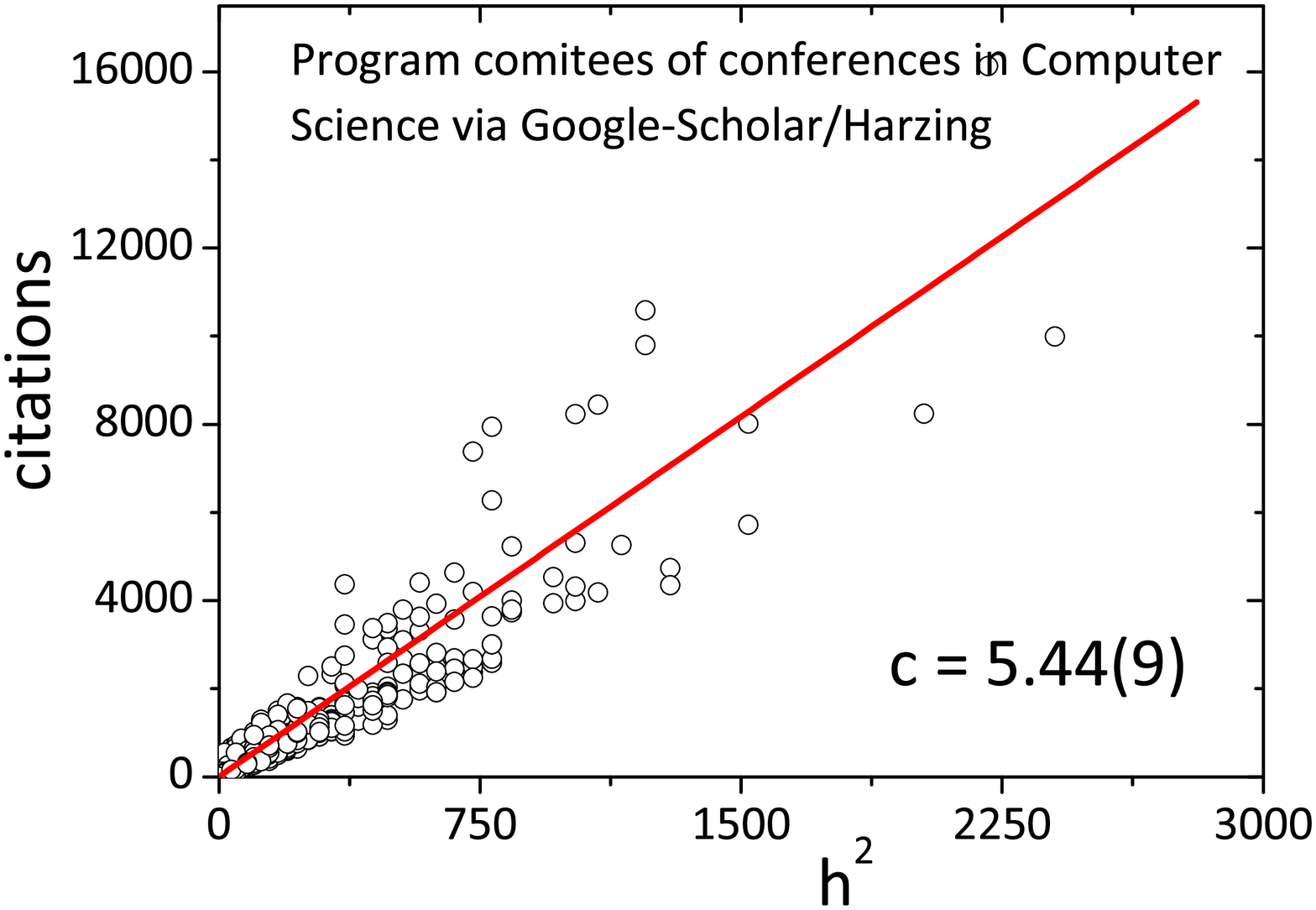} 
\\[0pt]
\textbf{(b)}
\end{center}
\caption{These figures exhibit the number of citations as function of $h^{2}$
of different authors, for different groups, used to estimate $c$ of Hirsch's
relation numerically obtained by \protect\ref{slope}. Plot (a), for members
of post-graduate programs from Brazil obtained via Lattes platform-ISI-JCR.
Plot (b), corresponds to data from members of committee programs of computer
science conferences obtained by google-Harzing.}
\label{fig:estimate_of_c}
\end{figure}

Our main results are presented below. Consider the researchers from Post-Graduations in Physics and Biology as Group I and from Conferences Computer Science (Harzing/Google Scholar) as Group II. 
The plots of citation number as a function of each author $h^2$ are depicted in Fig.~\ref{fig:estimate_of_c}. 
The plots in Fig. \ref{fig:estimate_of_c} exhibit the behavior of the citation number as a function of $h^{2}$, for different authors in each one of the studied groups. The proportionality parameter $c$, given by Eq.~\ref{slope}, is numerically estimated. 
Plot (a) corresponds to data from Group I ($c = 3.75(4)$) and Plot (b) is a similar plot for the Group II ($c = 5.44(9) $). These $\widehat{c}$ estimates reflect the difference between the two areas.

In what follows, we compute the parameters of citation pdf's of both
considered groups. The parameters of the stretched exponential pdf~\cite%
{Laherrere1998} and the generalized exponential pdf~\cite{Tsallis2000} are
estimated using the method of moments.

\subsection{Stretched Exponential PDF}

For accurately estimating $\beta $ in Eq.~(\ref{laeherre}), we firstly use
the method of moments and compare it with the value obtained from the Zipf
plot. The method of moments consists in calculating the $k^{\mbox{th}}$
moment of the pdf comparing them with the experimental ones. Consider the
moments of the stretched exponential pdf: 
\begin{equation*}
\left\langle x^{k}\right\rangle =\frac{\beta }{x_{0}\Gamma (1/\beta )}
\;\int_{0}^{\infty }dxx^{k}\;e^{-(x/x_{0})^{\beta }}=\frac{x_{0}^{k}}{\Gamma
(1/\beta )}\Gamma (\frac{k+1}{\beta }) \; .
\end{equation*}

The experimental moments are calculated as $\overline{x^{k}}%
=(x_{1}^{k}+x_{2}^{k}+...+x_{n}^{k})/n$. The method consists in calculating
the best $\beta$ value that matches both numerically calculated $%
\left\langle x^{k}\right\rangle $ and $\overline{x^{k}}$ for several $k$
values (not only for integers). Since $\left\langle x^{k}\right\rangle $
depends on $x_{0}$, one considers the ratio 
\begin{equation}
\Phi _{k}^{(\beta )}=\frac{\langle x^{k} \rangle}{\langle x \rangle^{k}} = 
\frac{\Gamma^{k-1}(1/\beta)\Gamma[(k+1)/\beta]}{\Gamma^{k}(2/\beta)}
\label{theoretical_sornette}
\end{equation}
to eliminate the $x_{0}$ dependence. The corresponding experimental ratio 
\begin{equation}
\Phi _{k}^{(\exp )}= \frac{\overline{x^{k}}}{\overline{x}^{k}}=\frac{%
\sum\nolimits_{i=1}^{n}x_{i}^{k}}{(\sum_{i=1}^{n}x_{i})^{k}} \; .
\label{experimental_sornette}
\end{equation}

Form the numerical minimization of $\int (\Phi _{k}^{(\beta
)}-\Phi_{k}^{(\exp )})^{2}dk$, using $\delta k=0.01$, one obtains $\beta
=0.47$ for Group I and $\beta =0.31$ for Group II. In Fig.~\ref%
{moments_sornette}, we show plots of theoretical moments for several $\beta$
values and the experimental moments in same plot for both groups studied
here.

\begin{figure}[h]
\begin{center}
\includegraphics[width=.8%
\columnwidth]{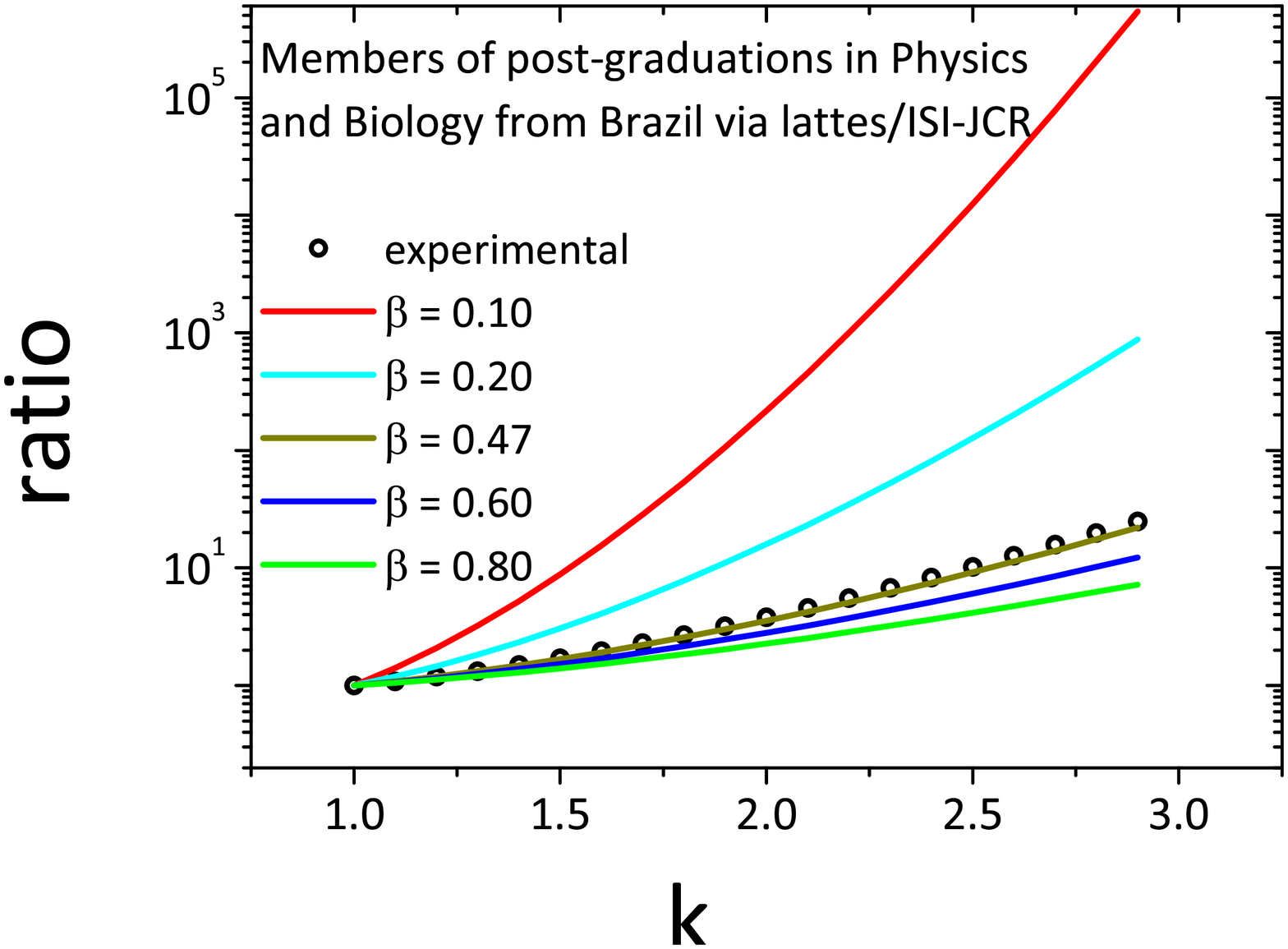} \\[0pt]
\textbf{(a)} \\[0pt]
\includegraphics[width=.8\columnwidth]{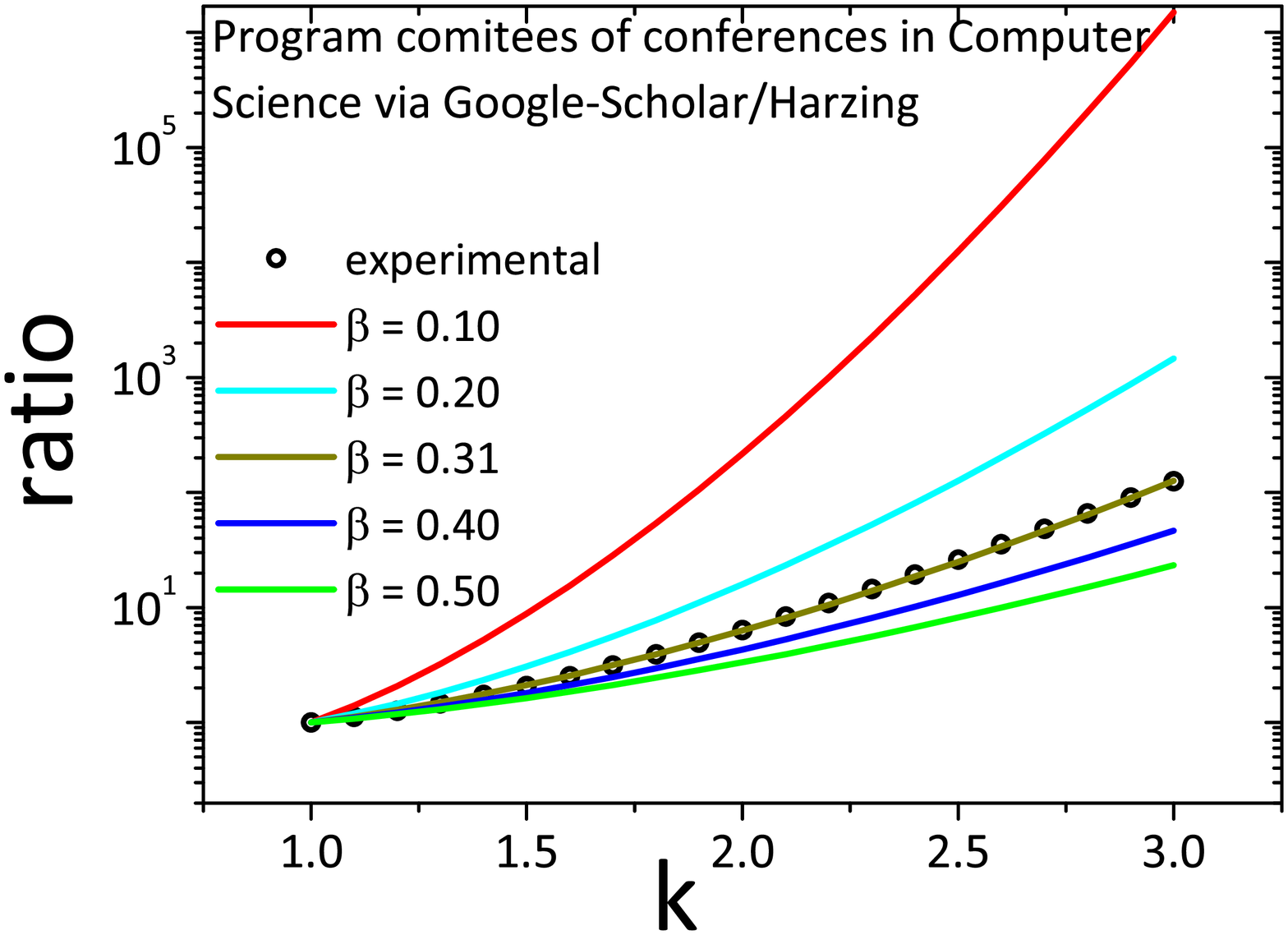} \\[0pt]
\textbf{(b)}
\end{center}
\caption{These plots exhibit the theoretical moments based on citations
stretched exponential pdf (Eq.~\protect\ref{theoretical_sornette}), compared
with experimental moments (Eq.~\protect\ref{experimental_sornette}) for
group I \textbf{(a)} with $\protect\beta =0.47$ and II \textbf{(b)} with $%
\protect\beta =0.31$.}
\label{moments_sornette}
\end{figure}

The result for Group I, even for the same database (ISI-JCR), clearly is
different from that found in Ref.~\cite{Laherrere1998}. However, the result
for Group I is closer of exponent for the citations of papers (not of
authors) from journal Physical Review D ($\beta \approx 0.39$) and similar
to citations of the papers from ISI ($\beta \approx 0.44$) obtained in Ref.~%
\cite{Redner1998}, in the limit of low citations ($x<500$). Although the
value found for Group II corroborates the value found in Ref.~\cite%
{Laherrere1998} ($\beta \approx 0.3$), no matching was expected because this
last result was obtained for citations of 1120 most cited authors obtained
from ISI-JCR, during a time-lag (between 1981-1997) and our results are
based on all citations of scientific life of considered authors.

To test the quality of our fits, we also considered suitable Zipf plots. The
main idea of Zipf plot is to rank the citations of all authors according to $%
x_{1}\geq x_{2}\geq x_{3}...\geq x_{n}$. The upper tail distribution is
expected to be: 
\begin{equation}
\zeta _{j}=\int_{x_{j}}^{\infty }P_{\beta }(x)dx=1-\frac{j}{n}
\label{zipf_relation}
\end{equation}%
where $j$ is the rank of citation $x_{j}$. For the stretched exponential,
one has: 
\begin{equation}
\zeta _{j}=\frac{\beta }{\widehat{x}}\frac{\Gamma (2/\beta )}{\Gamma
(1/\beta )}\int_{x_{j}}^{\infty }dx\exp [\frac{-\Gamma (2/\beta )^{\beta }}{%
\Gamma (1/\beta )^{\beta }\overline{x}^{\beta }}x^{\beta }]=\frac{\Gamma
(1/\beta ,\frac{\Gamma (2/\beta )^{\beta }x_{j}^{\beta }}{\Gamma (1/\beta
)^{\beta }\widehat{x}^{\beta }})}{\Gamma (1/\beta )} \; ,
\label{zipf_stretched_equation}
\end{equation}%
where $\Gamma (a,b)=\int_{b}^{\infty }z^{a-1}e^{-z}dz$ is known as the
incomplete gamma function.

\begin{figure}[h]
\begin{center}
\includegraphics[width=.8%
\columnwidth]{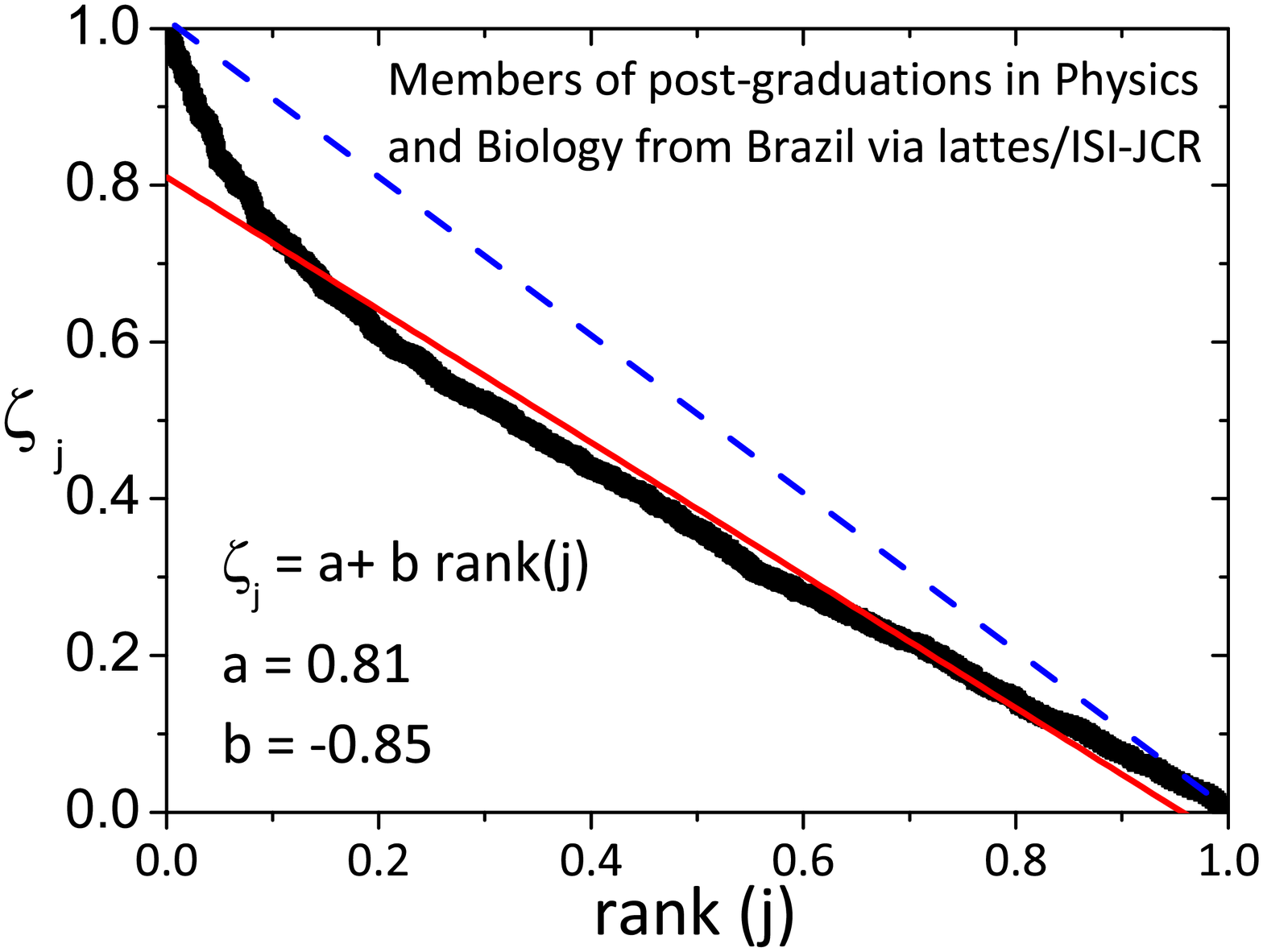} \\[0pt]
\textbf{(a)} \\[0pt]
\includegraphics[width=.8\columnwidth]{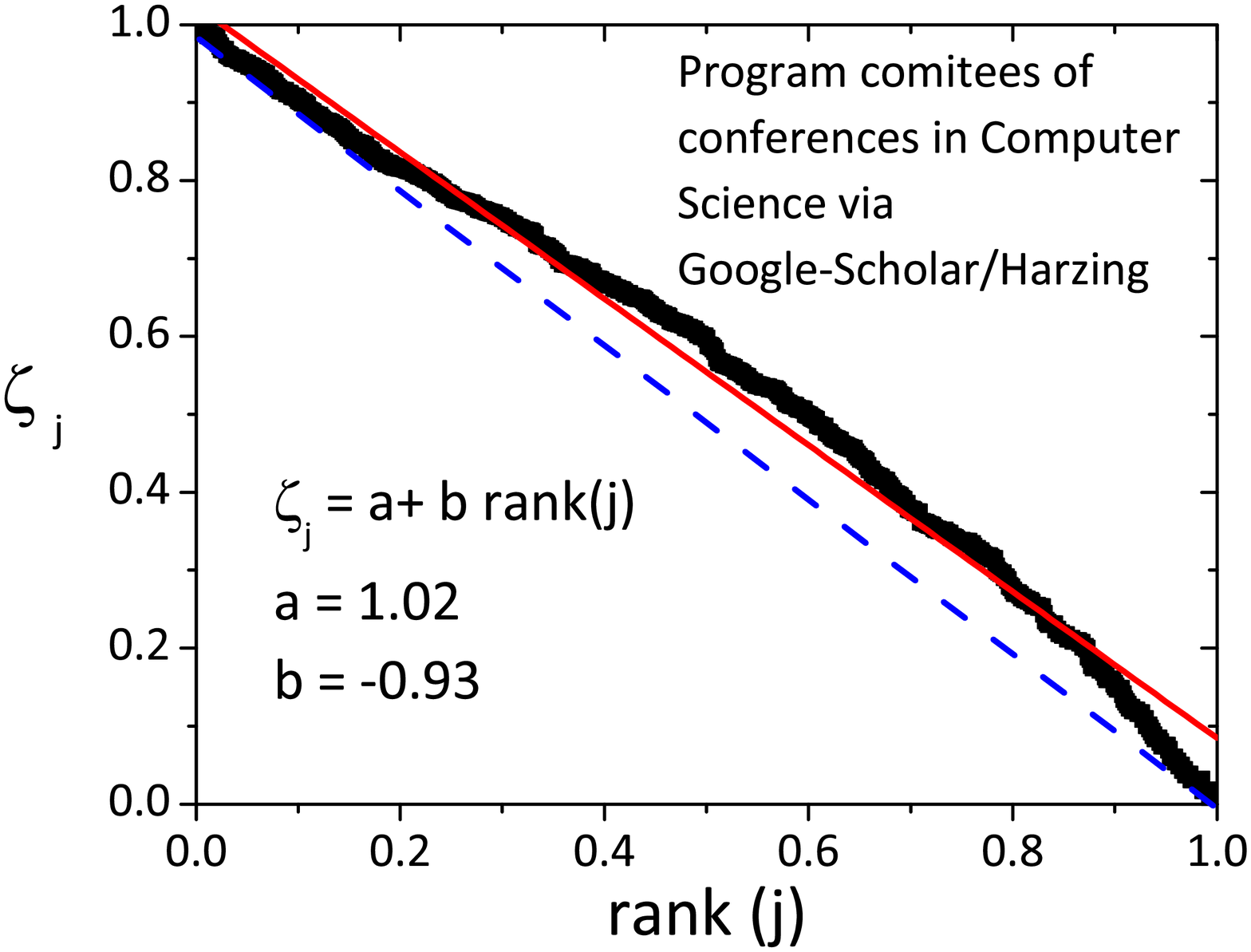} 
\\[0pt]
\textbf{(b)}
\end{center}
\caption{Zipf plots for the stretched exponential pdf for: \textbf{(a)}
Group I and \textbf{(b)} Group II. The expected linear behavior by relation 
\protect\ref{zipf_relation} is tested plotting $\protect\zeta _{j}$
calculated by \protect\ref{zipf_stretched_equation} as function of $j/n$. }
\label{zipf_stretched}
\end{figure}

In Fig~\ref{zipf_stretched}, we display the Zipf plots ($\zeta _{j}$ as
function of $j/n$), using $\beta =0.47$ for Group I and $\beta =0.31$ for
Group II. One can see a near linear behavior with slope close to $-1$ and
intercept close to $1$, the expected values for both cases. However, some
discrepancies are present. For sake of comparison, we show the linear fit
(red continuous line) and an exact expected behavior (dashed blue line).

\subsection{Generalized Exponential PDF}

\begin{figure}[h]
\begin{center}
\includegraphics[width=.8%
\columnwidth]{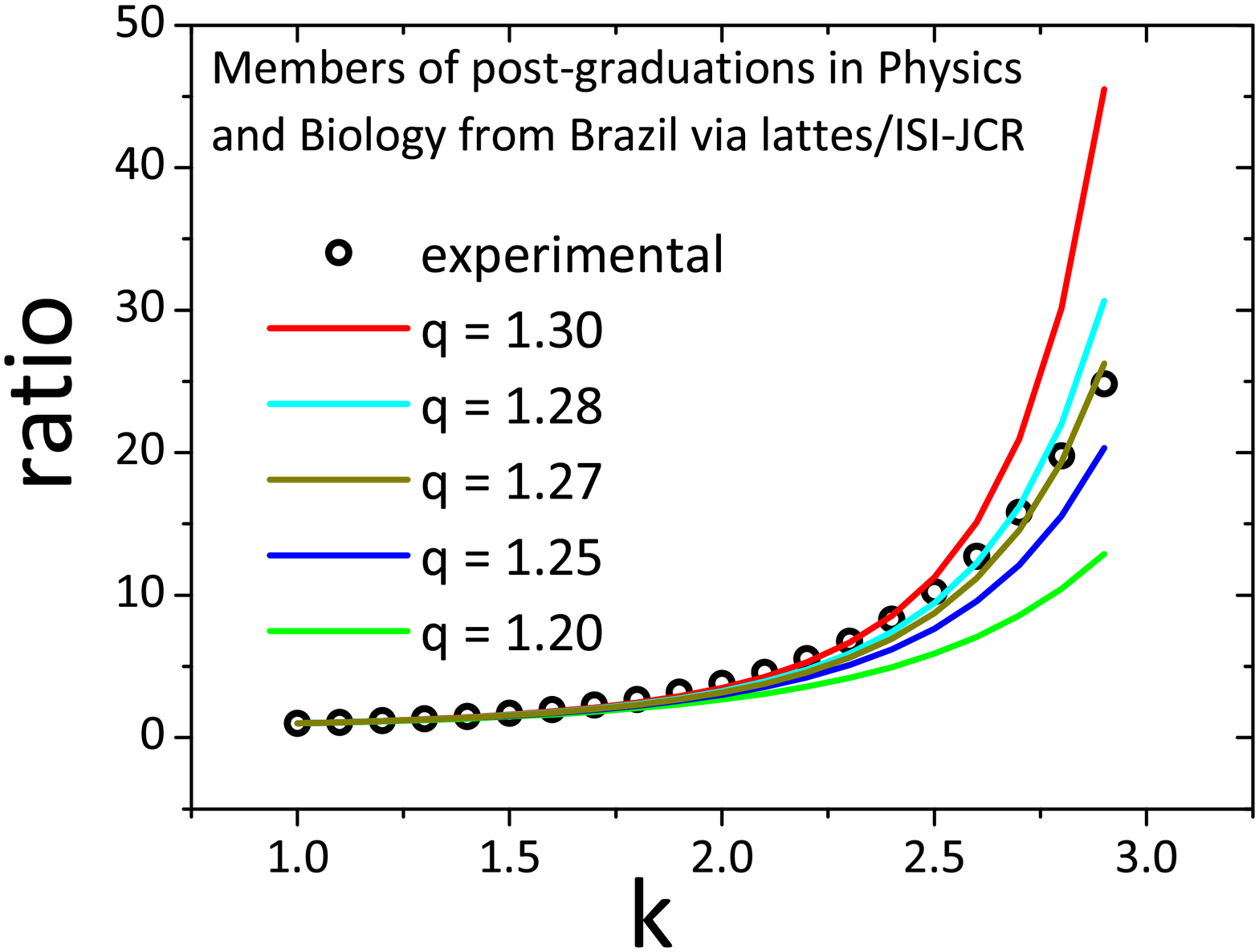} \\[0pt]
\textbf{(a)} \\[0pt]
\includegraphics[width=.8\columnwidth]{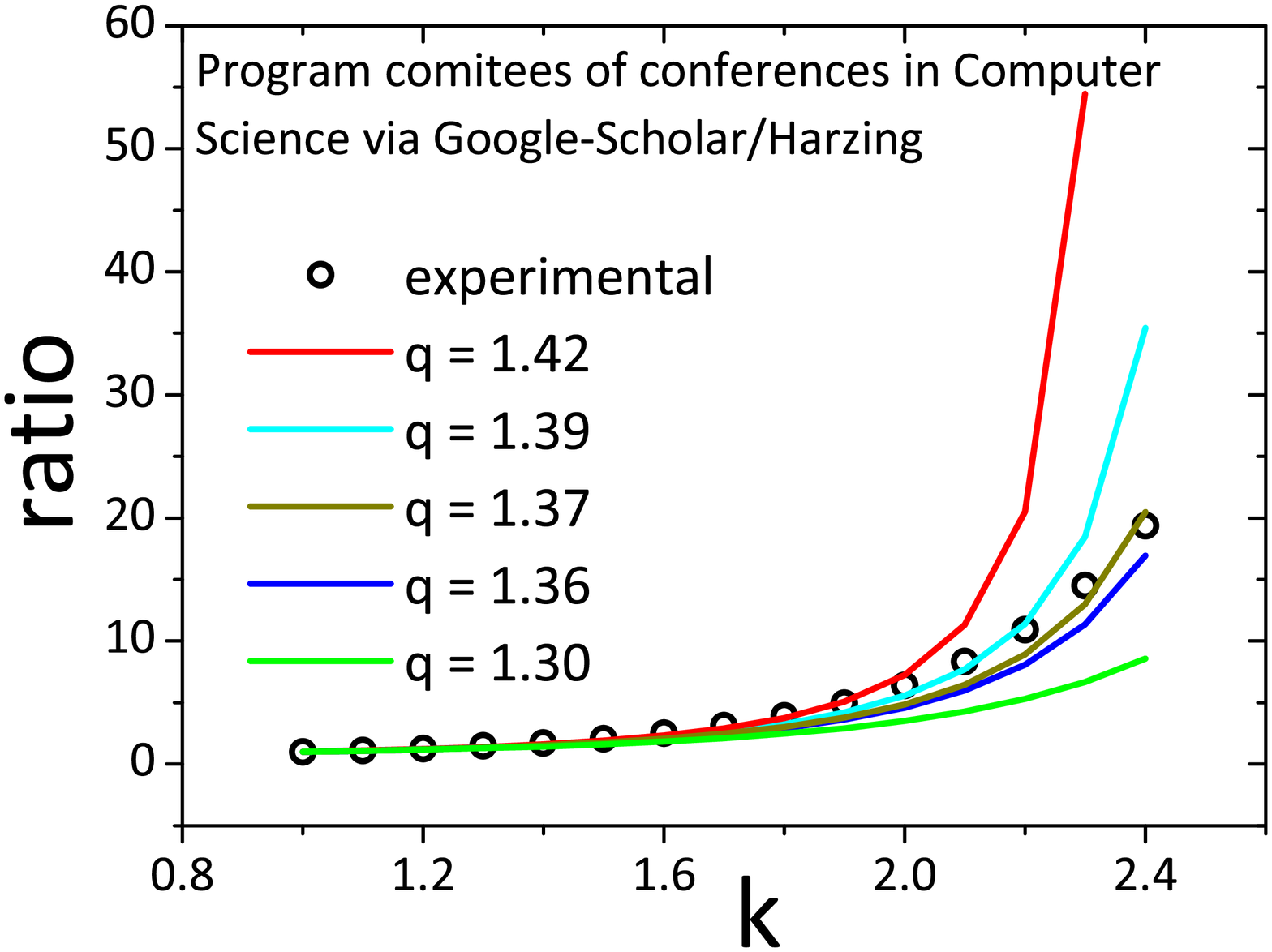} \\[0pt]
\textbf{(b)}
\end{center}
\caption{ These plots exhibit the theoretical moments based on citations
generalized exponential pdf (Eq.~\protect\ref{theoretical_generalized}),
compared with experimental moments (Eq.~\protect\ref{experimental_sornette})
for group I \textbf{(a)} with $q=1.27$ and II \textbf{(b)} with $q=1.37$. }
\label{generalized_moments}
\end{figure}

Now we consider the generalized exponential pdf. The moment method is
addressed using the ratio: 
\begin{equation}
\Psi _{k}=\frac{\left\langle x^{k}\right\rangle _{q}}{\left\langle
x\right\rangle _{q}^{k}}=\frac{1}{(2-q)\widehat{x}^{k+1}}\int_{0}^{\infty
}dx\ x^{k}\left[ 1+\frac{(q-1)}{(2-q)\overline{x}}x\right] ^{q/(1-q)} \; ,
\label{theoretical_generalized}
\end{equation}%
where only the first moment $\left\langle x\right\rangle _{q}$ is estimated
as simple averages: $\widehat{x}=473(23)$ for Group I and $\widehat{x}%
=936(76)$, for Group II. Similarly to the stretched exponential case, we use
the same procedure: compare $\Psi _{k}$ with the experimental moments ($%
\Phi_{k}^{(\exp )}$) calculated by Eq.~(\ref{experimental_sornette}). These
plots are displayed in Fig.~\ref{generalized_moments} and the best adjusted
values are $q=1.27$, for Group I and $q=1.37$, for Group II.

\begin{figure}[h]
\begin{center}
\includegraphics[width=.8%
\columnwidth]{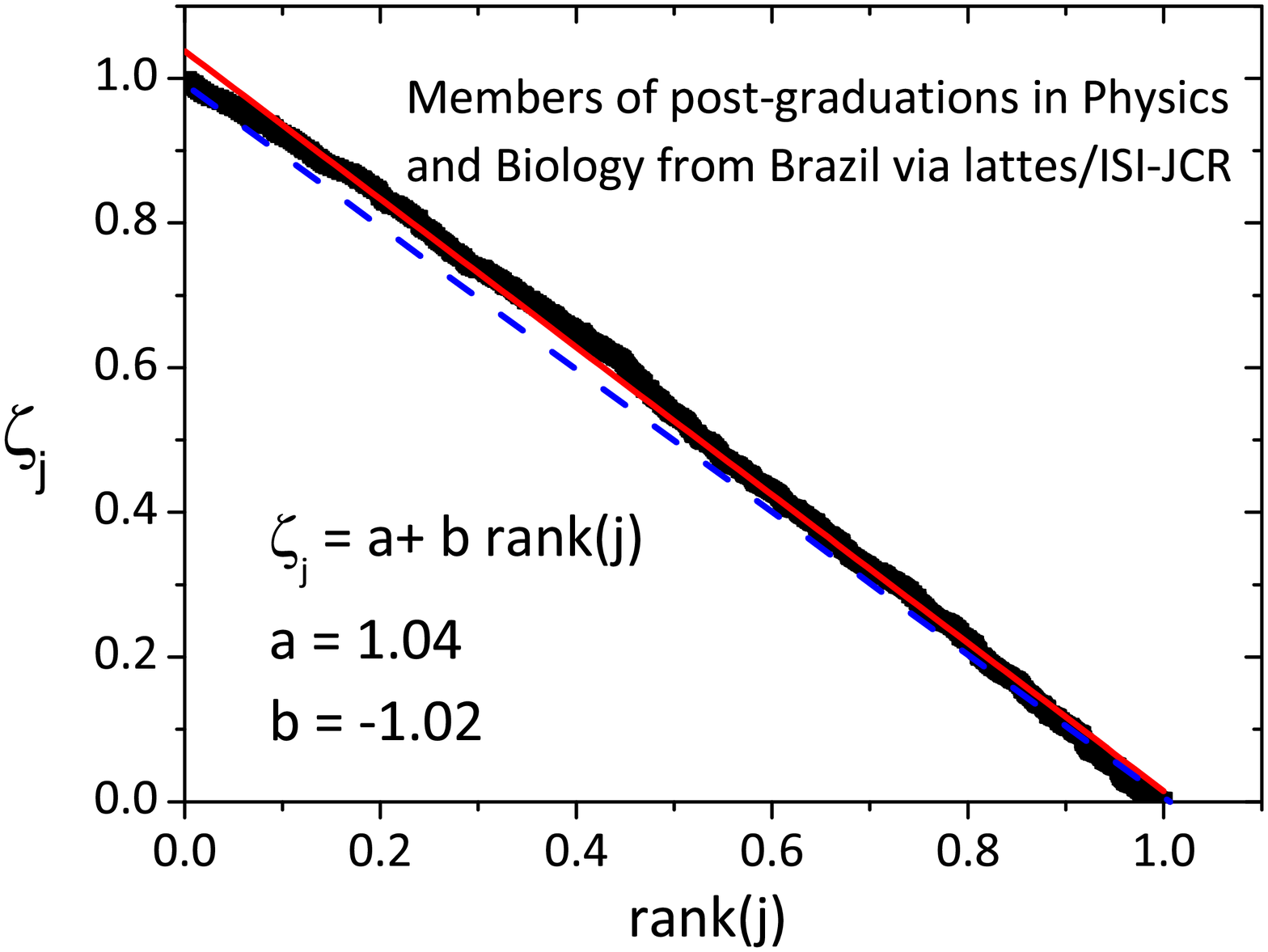} \\[0pt]
\textbf{(a)} \\[0pt]
\includegraphics[width=.8\columnwidth]{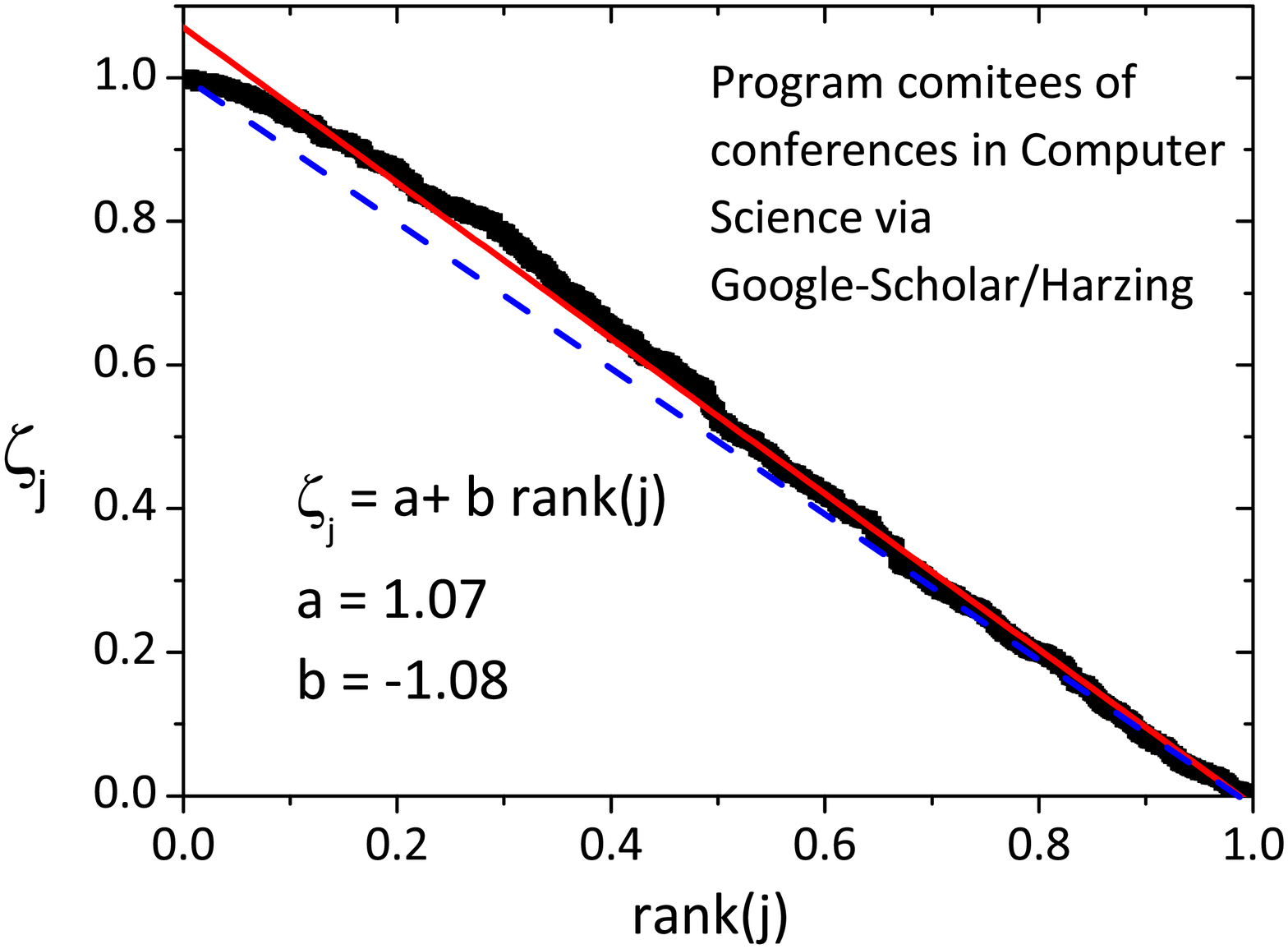}
\\[0pt]
\textbf{(b)}
\end{center}
\caption{ Zipf plots for the stretched exponential pdf for: \textbf{(a)}
Group I and \textbf{(b)} Group II. }
\label{zipf_generalized}
\end{figure}

The upper tail distribution for the generalized exponential is:%
\begin{equation*}
\begin{array}{lll}
\zeta _{j} & = & \int_{x_{j}}^{\infty }P_{q}(x)dx \\ 
&  &  \\ 
& = & \frac{1}{(2-q)\overline{x}}\int_{1+\frac{(q-1)}{(2-q)\overline{x}}%
x_{j}}^{\infty }x^{q/(1-q)}dx \\ 
&  &  \\ 
& = & \exp _{q}\left( \frac{-x_{j}}{(2-q)\overline{x}}\right)%
\end{array}
\; .
\end{equation*}
Similarly to Fig.~\ref{zipf_stretched}, we have used the values of $\widehat{%
x}$ to plot $\zeta _{j}$ as function of rank ($j/n$) as illustrated in Fig.~%
\ref{zipf_generalized}.

From Figs.~~\ref{zipf_stretched} and~\ref{zipf_generalized}, one observes a
better linear fit for the Zipf plots using the generalized exponential pdf.
However an important question is if the same values of $q$ here estimated
for citation distribution are also estimated when we perform $h$-index
distribution fits.

\subsection{$h$ index PDF}

In Table~\ref{parameters}, we summarize the estimated parameters for the
stretched and generalized exponential pdf's using the method of moments and
Zipf plots.

\begin{table}[tbp]
\centering%
\begin{tabular}{|l|c|c|c|c|c|}
\hline\hline
\textbf{Groups} & $c$ & $\widehat{c}$ & $\beta $ & $\widehat{x}$ & $q$ \\ 
\hline\hline
I & 3.75(4) & $3.77(5)$ & 0.47(1) & 473(23) & 1.27(1) \\ \hline
II & 5.44(9) & $5.71(11)$ & 0.31(1) & 936(76) & 1.37(1) \\ \hline\hline
\end{tabular}%
\caption{Summary of the parameter estimated by the method of moments for the
citation distribution pdf which was fitted as stretched exponential and as
generalized exponential. Group I refers Researchers from Post-Graduations in
Physics and Biology (Lattes/CNPq-ISI-JCR). Group II refers Conferences
Computer Science (Harzing/Google Scholar). The coefficient $c$ is obtained
by fits of the plots of Fig.~\protect\ref{fig:estimate_of_c}. The
coefficient estimator $\widehat{c}$ is calculated from Eq.~\protect\ref%
{slope}. The stretched exponential parameter $\protect\beta $ is obtained by
the best matching via method of moments shown in Fig.~\protect\ref%
{moments_sornette}. The average estimator $\widehat{x}$ is the simple
arithmetic mean. The generalized exponential parameter $q$ is similarly
obtained according to Fig.~\protect\ref{generalized_moments}. }
\label{parameters}
\end{table}

\begin{figure}[h]
\begin{center}
\includegraphics[width=.8\columnwidth]{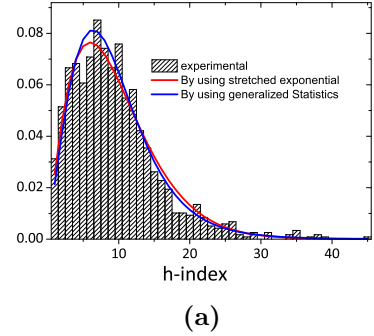} \\[%
0pt]
\textbf{(a)} \\[0pt]
\includegraphics[width=.8\columnwidth]{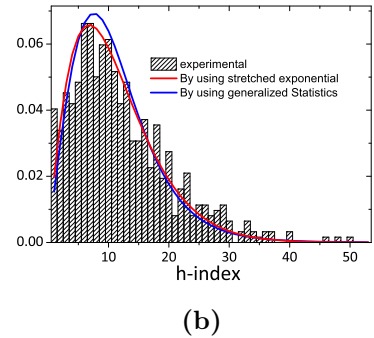} \\[0pt]
\textbf{(b)}
\end{center}
\caption{ $h$-index distribution for \textbf{(a)} Group I and \textbf{(b)}
Group II. Blue line corresponds to the stretched exponential pdf (Eq.~%
\protect\ref{h_distribution_by_sornete}) and the red line to the generalized
exponential pdf (Eq.~\protect\ref{h-index_distribution}). }
\label{h_distribution}
\end{figure}

Let us now consider the databases $h$-indices pdfs compared to the stretched
and generalized exponential pdf's [Eqs.~(\ref{h_distribution_by_sornete})
and~(\ref{h-index_distribution})]. Using the estimates of $\widehat{c}$ and $%
\widehat{x}$ for each group studied (see table \ref{parameters}), we find
numerically the $\beta $ that minimizes $\chi _{\beta
}^{2}=\sum_{h=h\min}^{h_{\max }}\left[ f^{(\exp )}(h)-H_{\beta }(h)\right]
^{2}$ for the stretched exponential pdf and $q$ that minimizes $%
\chi_{q}^{2}=\sum_{h=h\min }^{h_{\max }}\left[ f^{(\exp )}(h)-H_{q}(h)\right]%
^{2}$ for the generalized exponential pdf, where $H_{\beta }(h)$ is computed
by Eq~(\ref{h_distribution_by_sornete}) and $H_{q}(h)$ by Eq.~(\ref%
{h-index_distribution}). Our computer code runs with $q$ ranging from $%
q_{\min }=1.01$ up to a $q_{\max }=1.99$ and $\beta$, from $\beta
_{\min}=0.01$ up to a $\beta _{\max }$ $=0.99$ in steps of $\Delta q=\Delta
\beta =0.01$. The best fits, according to the $\chi ^{2}$ measures, give $%
\beta =0.66(1)$ and $q=1.26(1)$, for Group I and $\beta =0.64(1)$ and $%
q=1.24(1)$, for Group II using the stretched and generalized exponential
pdf's, respectively. Both pdf's are depicted in Fig.~\ref{h_distribution}
and the comparison among the estimated parameters to the ones of Table~\ref%
{parameters}, is compiled in Table~\ref{obtained_parameters}.

\begin{table}[tbp]
\centering%
\begin{tabular}{|l|c|c|c|c|}
\hline\hline
\textbf{Groups} & $\beta _{m}$ & $\beta _{\chi ^{2}}$ & $q_{m}$ & $q_{\chi
^{2}}$ \\ \hline\hline
I & 0.47(1) & 0.66(1) & 1.27(1) & 1.26(1) \\ \hline
II & 0.31(1) & 0.64(1) & 1.37(1) & 1.24(1) \\ \hline\hline
\end{tabular}%
\caption{Comparison of the stretched and generalized exponential pdf's
parameters estimated by the method of moments and $\protect\chi ^{2}$
procedures. One sees that the generalized exponential pdf, has a more robust
estimation that the stretched exponential pdf.}
\label{obtained_parameters}
\end{table}

Although, from Fig.~\ref{h_distribution}, one can see good fits for both
pdf's, from Table~\ref{obtained_parameters}, one observes that generalized
exponential pdf produces a much better matching among the estimated
parameter, via citation distribution through the moment method and $h$-index
distribution through the $\chi^2$ method than the stretched exponential pdf.
For Group I, for the generalized exponential pdf, we have an exact estimated
parameter showing its greater robustness when compared to the stretched
exponential pdf. It is important to notice that for Group II, both pdf's
produce more distant estimates.

Our results indicate that the generalized exponential pdf is more
appropriate to describe the $h$-index pdf, supplying an interesting and
simple formulae for $h$-indices of very distinct groups in different
databases. Since the data have been collected from very different sources,
one can claim the universal aspect of the generalized exponential pdf to
represent the continuous $h$-index.

\section{Conclusions}

\label{conclusions}\textbf{\ }

In the first part of this manuscript, we analyze the different formulas for citation distribution calculating their parameters via two different methods: method of moments and by Zipf plots for two very distinct groups of researchers pertaining to different databases. 
In the second part, we calculate the $h$-index distribution also for these different databases to find a universal formula.
Our results show that good fits can be obtained for the $h$-index pdf using suitable estimates and the relation $x=ch^{2}$. 
It is also important to mention that we have estimated the parameters $\beta$ and $q$ in two independent ways and by moments and $\chi^2$ methods: the citation distribution of Eqs.~(\ref{laeherre}) and~(\ref{citation_generalized}) and $h$-index distribution of Eqs.~(\ref{h_distribution_by_sornete}) and~(\ref{h-index_distribution}). 
Such fits produce more similar results for the $q$ distributions.  
%

\section*{Acknowledgments}

R. da Silva (308750/2009-8), A.S. Martinez (305738/2010-0 and 476722/2010-1)
and, J.P. de Oliveira (476722/2010-1) are partly supported by the Brazilian
Research Council CNPq.


\end{document}